\documentclass[aps,prl,twocolumn,groupedaddress]{revtex4}
\usepackage{graphicx}
\usepackage{dcolumn}
\usepackage{bm}
\begin{document}

\title{Isotope effects in  high-T$_c$ cuprate superconductors:
Ultimate proof for  bipolaron theory of superconductivity}

\author{A. S. Alexandrov}
\affiliation{Department of Physics, Loughborough University, Loughborough, LE11 3TU, U.K.}

\author{G. M. Zhao}
\affiliation{Department of Physics and Astronomy, California State University, Los Angeles, CA 90032, USA}

\begin{abstract}

\textbf{Developing a theory of high-temperature superconductivity in
copper oxides is one of the outstanding problems in physics.
 Twenty-five years after its discovery \cite{bedmul}, no consensus on the
 microscopic theory has been reached
despite tremendous theoretical and experimental efforts. Attempts to
understand this problem are hindered by the subtle interplay among a
few mechanisms and the presence of several nearly degenerate and
competing phases in these systems. Here we provide unified
parameter-free explanation of the observed oxygen-isotope effects on
the critical temperature, the magnetic-field penetration depth, and
on the normal-state pseudogap for underdoped cuprate superconductors
within the framework of the bipolaron theory compatible with the
strong Coulomb and Fr\"ohlich interactions, and with many other
independent observations in these highly polarizable doped
insulators. Remarkably,  we also quantitatively explain the measured
critical temperature and the magnitude of the magnetic-field
penetration depth. The present work thus represents an ultimate
proof of the bipolaron theory of high-temperature superconductivity,
which takes into account essential Coulomb and electron-phonon
interactions. } \pacs{71.38.-k, 74.40.+k, 72.15.Jf, 74.72.-h,
74.25.Fy}

\end{abstract}

\maketitle

In 1911 curiosity concerning the electrical properties of metals at
low temperatures led the Dutch physicists,  Kamerling Onnes and his
assistant G. Holst to discover superconductivity at 4.2 K in mercury
\cite{kam}. This discovery was one of the most important
experimental findings in low temperature physics.  On the long way towards a microscopic
understanding of superconductivity, the observation of an isotope
effect on the critical temperature, $T_{c}$ in 1950 \cite{max,rey}
gave an important clue to the microscopic
mechanism of superconductivity. The presence of an isotope
effect thus implies that superconductivity is not of purely
electronic origin. In the same year  Fr\"ohlich \cite{fro} pointed
out that the electron-phonon interaction gave rise to an attractive
interaction between electrons, which might be responsible for
superconductivity. Fr\"ohlich's theory played a decisive role in
establishing the correct mechanism. Finally, in 1957, Bardeen,
Cooper and Schrieffer \cite{bar} (BCS) developed the BCS theory that
was the first successful microscopic theory of superconductivity.
The BCS theory implies an isotope-mass dependence of $T_{c}$, with
an isotope-effect exponent $\alpha=-d\ln T_c/d\ln M = 1/2$, in
excellent agreement with the reported isotope exponents in  simple
metallic superconductors like Hg, Sn and Pb.

The doping dependent oxygen-isotope effect (OIE) on the critical temperature
$T_c$, $\alpha^{O}=-d\ln T_c/d\ln M_{O}$  (where $M_{O}$ is the
oxygen-isotope mass)\cite{polarons} and the substantial OIE on the
in-plane supercarrier mass $m_{ab}^{**}$, $\alpha^{O}_{m^*} =
dm_{ab}^{**}/d \ln M_{O}$ (ref.~\cite{ZhaoYBCO95,ZhaoLSCO95,ZhaoNature97,ZhaoJPCM98,Zhao01,Keller1}), provide direct evidence for a
significant electron-phonon interaction (EPI) also in
high-temperature cuprate superconductors. High resolution
angle-resolved photoemission spectroscopy (ARPES) \cite{lanzara}
provides further evidence for the strong EPI  apparently with
c-axis-polarised optical phonons. These results along with optical
\cite{mic}, neutron scattering \cite{ega,rez}, and tunneling
 data \cite{Zhao07,Boz08,ZhaoPRL09} unambiguously show that lattice vibrations  play a
significant but unconventional role in high-temperature
superconductivity. The interpretation of the optical spectra of
high-T$_c$ materials as due to multi-polaron absorption \cite{dev}
strengthens the view \cite{asa} that the Fr\"{o}hlich EPI is
important in those structures. Operating together with a
shorter-range deformation potential and molecular-type (e.g.,
Jahn-Teller \cite{muller:2000}) EPIs, the Fr\"ohlich EPI can readily
overcome the Coulomb repulsion at a short distance of about the
lattice constant for electrons to form  real-space intersite
bipolarons (see \cite{alebook} and Supplementary Information \cite{sup}).

Despite all these remarkable and well-done experiments that lead to
the consistent conclusion about the important role of EPI in
high-temperature superconductors, there is no consensus on the
microscopic origin of the observed unconventional isotope effects on
the in-plane magnetic-field penetration depth and the normal-state
pseudogap. The doping dependent $\alpha^{O}$ has been explained as due
to the doping independent oxygen-isotope effect on the in-plane
carrier concentration $n$, that is, $\alpha^{O}_{n}$ = $-d\ln n/d\ln
M_{O}$
= 0.146 (ref.~\cite{Muller2011}). This interpretation contradicts
other independent experiments \cite{ZhaoNature97,ZhaoJPCM98,Keller1}
which consistently show that the carrier concentrations of the two
oxygen-isotope samples are the same within 0.0004 per Cu. This is
also in sharp contrast to the observed very large oxygen-isotope
effect on the low-temperature magnetic-field penetration depth, $\lambda_{ab}\propto (m_{ab}^{**}/n)^{1/2}$, in both
La$_{1.94}$Sr$_{0.06}$CuO$_{4}$ (ref.~\cite{ZhaoJPCM98}) and
Y$_{0.55}$Pr$_{0.45}$Ba$_{2}$Cu$_{3}$O$_{7-y}$
(ref.~\cite{Keller2}), which would lead to $\alpha^{O}_{n}$ $\simeq$ 2
if one would assume that the supercarrier mass is independent of the
oxygen mass. Another model based on the pair-breaking effects due to
impurities, disorder, and/or pseudogap can also explain the observed
oxygen-isotope effects on the penetration depth and the critical
temperature in deeply underdoped samples \cite{Tallon}. But this
model cannot consistently explain the negligibly small $\alpha^{O}$ but
the large OIE on the penetration depth in optimally doped samples
\cite{Zhao01}.

Alternatively, the bipolaron theory of superconductivity
\cite{alebook} can naturally account for the substantial
$\alpha^{O}_{m^*}$ and large $\alpha^{O}$ in deeply underdoped cuprates
\cite{aleiso}. There is a qualitative difference between ordinary
metals and polaronic conductors. The renormalized effective mass
of electrons is independent of the ion mass $M$ in ordinary metals
(where the Migdal adiabatic approximation is believed to be valid),
because the EPI coupling constant $\lambda$ does not depend on the
isotope mass (see, for instance ref.~\cite{sup}). However, when
electrons form polarons (new quasiparticles dressed by lattice
distortions), their effective mass $m^{*}$ depends on $M$
\cite{aleiso}.

Although the bipolaron theory can qualitatively explain both
$\alpha^{O}_{m^*}$ and $\alpha^{O}$ in deeply underdoped samples, some
important issues have not been well addressed by the theory. The
first issue is why $\alpha^{O}_{m^*}$ is not equal to $\alpha^{O}$ even in
deeply underdoped cuprates. The second issue is why $\alpha^{O}$ is much
smaller than $\alpha^{O}_{m^*}$ for slightly underdoped samples. The
third issue is why there is a giant negative oxygen-isotope effect
on the pesudogap formation temperature $T^{*}$ in
HoBa$_{2}$Cu$_{4}$O$_{8}$ (ref.~\cite{Temp}).  Here we provide
parameter-free explanations to the observed oxygen-isotope effects
on the critical temperature, the in-plane supercarrier mass, and on
the normal-state pseudogap in
HoBa$_{2}$Cu$_{4}$O$_{8}$ within the framework of the bipolaron
theory. The present work thus represents an ultimate proof of the
bipolaron theory of high-temperature superconductivity.

\begin{figure}
\begin{center}
\includegraphics[angle=-00,width=0.45\textwidth]{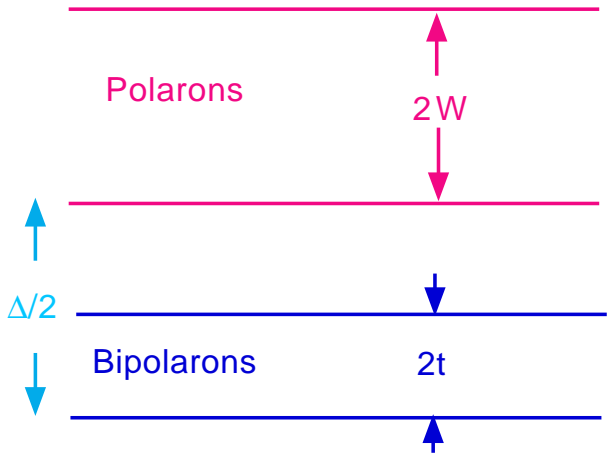}
\vskip -0.5mm \caption{(Color online) Low energy excitations in
cuprate superconductors.}
\end{center}
\end{figure}

We adopt the bipolaronic low-energy excitation structure of
cuprates, Fig.~1, derived from the microscopic Hamiltonian with the
strong Coulomb and Fr\"ohlich interactions \cite{alebook,sup}. The
critical temperature of quasi-two-dimensional (2D) bipolarons, which are hard-core
bosons, depends on their density $n_b(T_c)$ at the critical
temperature and
 the in-plane bipolaron
 mass $m^{**}_{ab}$:
\begin {equation}
T_c \propto
\frac{n_b(T_c)}{m^{**}_{ab}}.
\label{Tc}
\end{equation}

The  bipolaron density slightly depends on temperature due to
bipolaron depletion into unbound single polarons, $n_b(T)=
[x-n_p(T)]/2$, with the density $n_p$ given by

\begin{equation}
n_p(T)= \frac{k_B T}{W}\ln (1+e^{-\Delta/2k_BT}),
\label{density}
\end{equation}
where $W$ is the polaron half-bandwidth, $x$ is the in-plane doping
level, and $\Delta$ is the
bipolaron binding energy, Fig.~1. This expression is obtained by
integrating the Fermi-Dirac distribution function with a constant
(2D) density of states in the polaron band, $N(E)=1/2W $ and assuming
that the polaron half-bandwidth is large enough, $W  \gg k_BT_c$,
\begin{equation}
n_p(T)= \int_0^{2W} dE {2N(E)\over{1+\exp[(\Delta/2+E)/k_B
T]}}.\label {density2}
\end{equation}
Here we take into account that the chemical potential is zero in the
superconducting state at and below $T_c$ if all energies are taking
with respect to the bipolaron ground state. The polaron in-plane
effective mass, $m^{*}_{ab} \propto \exp(A\sqrt{M})$ ($A$ is a
constant),  the
polaron inverse bandwidth, $1/W $, and the inter-site bipolaron
 mass $m^{**}_{ab}$ have the same isotope exponent $d\ln
m^{**}_{ab}/ \ln M \equiv \alpha_{m^{*}}=(1/2) \ln(m^{*}_{ab}/m)$
\cite{aleiso}, where $m$ is the band mass in a rigid lattice.  The
isotope effect on the pseudogap is given by:
\begin{equation}
    \delta\Delta = -\frac{3}{2}\delta W  = \frac{3W}{2} \frac{\delta
    m^{**}_{ab}}{m^{**}_{ab}},
    \label{gapoie}
    \end{equation}
   and
\begin{equation}
\frac{d\ln \Delta}{d\ln M} = \alpha_{m^*}\frac{3W}{2\Delta}.
\end{equation}

The above expressions are obtained by taking into account that the
pseudogap in Fig.~1 is given by $\Delta/2=J_p/2 -(W -t /2) =J_p/2 -
3W/4$, where $J_p$ is the phonon-induced intersite attraction, which is
\emph{independent} of the ionic mass \cite{sup}, and $t \approx W/2$ in
the intermediate coupling regime \cite{hague}. Then using Eqs.
(\ref{Tc}, \ref{density}) and neglecting the terms on the order of
$k_BT_c/W \ll 1$, one readily obtains the ratio:
\begin{equation}
\frac{\alpha}{\alpha_{m^*}}=1- \frac{1}{[x-n_{p}(T_{c})][1+\exp(\Delta/2k_B
T_c)]}.
\label{ratio}
\end{equation}

Eq.~\ref{ratio} can naturally explain why $\alpha^{O}_{m^*}$ is always
larger than $\alpha^{O}$ \cite{ZhaoNature97,ZhaoJPCM98}. It is worth noting that Eq.~\ref{ratio} is valid only if $d\ln
\Delta/d \ln M$ is small. When $d\ln \Delta/d \ln M$ is large, we
need to use Eqs.~\ref{Tc} and ~\ref{density} to directly calculate
the $T_{c}$ and $n_{p}(T_{c})$ changes upon the isotope exchange,
that is,

\begin{equation}
    \frac{\delta T_{c}}{T_{c}} = -\frac{\delta
    m^{**}_{ab}}{m^{**}_{ab}} - \frac{\delta n_{p}(T_{c})}{x-n_{p}(T_{c})}
    \label{tcoie}
    \end{equation}

Using the above equations we can quantitatively explain the  oxygen-isotope
effects on the pesudogap and the critical
temperature in slightly underdoped HoBa$_{2}$Cu$_{4}$O$_{8}$
(ref.~\cite{Temp}). The oxygen
isotope effect on the relaxation rate of crystal-field excitations in
this compound was investigated by means of inelastic neutron scattering
\cite{Temp}.
The relaxation rate, which is related to the free-carrier spin density,
clearly shows a large oxygen-isotope effect (see Fig.~2).
For the $^{16}$O sample there is evidence for the opening of an electronic gap in the normal
state at $T^{*}$ $\simeq$  170~K while for the $^{18}$O sample $T^{*}$ is shifted to about
220~K. In contrast, the $T_{c}$ is shifted from 79.0 to 78.5 K upon replacing $^{16}$O with $^{18}$O (the $^{18}$O
concentration is about 75$\%$) \cite{Temp}.

\begin{figure}
\begin{center}
\includegraphics[angle=-00,width=0.45\textwidth]{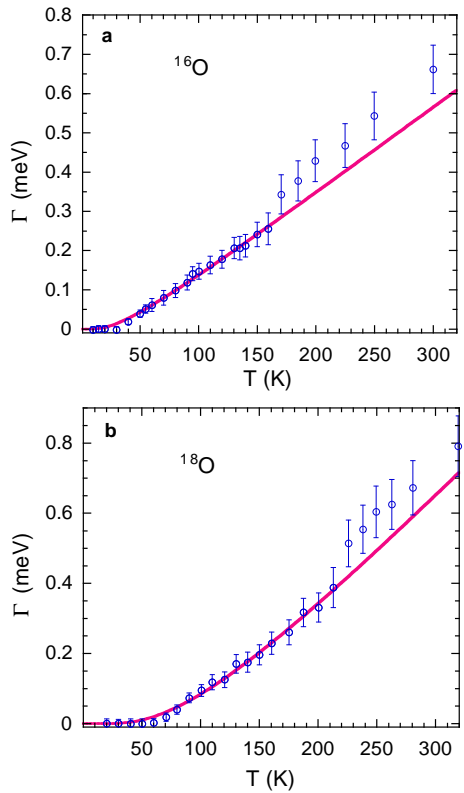}
\vskip -0.5mm \caption{(Color online) The relaxation rate $\Gamma$ of
crystal-field excitations for the  $^{16}$O and $^{18}$O samples of
slightly underdoped HoBa$_{2}$Cu$_{4}$O$_{8}$. The data are taken from
ref.~\cite{Temp}. The solid red lines represent the best fits of Eq.~\ref{rate} to the data
below $T^{*}$.
The best fits yield $\Delta/k_{B}$ = 94.6$\pm$8.8~K for the $^{16}$O
sample and $\Delta/k_{B}$ = 286$\pm$19~K for the $^{18}$O
sample.}
\end{center}
\end{figure}

In a normal state with no pseudogap, the relaxation rate $\Gamma_{n}(T)$ is proportional
to $[J_{ex}N(E_{F})]^{2}T$, where $J_{ex}$ is the exchange integral between the
4$f$ electrons of the Ho$^{3+}$ ions and the charge carriers and $N(E_{F})$ is the electronic
density of states at the Fermi energy \cite{Temp}. Within the polaron/bipolaron
framework, $N(E_{F})$ = $1/2W$ $\propto$ $m_{ab}^{*}$ $\propto$
$m_{ab}^{**}$, so the oxygen-isotope effect on $N(E_{F})$ is the same
as the oxygen isotope effect on $m_{ab}^{**}$. For a slightly underdoped
YBa$_{2}$Cu$_{3}$O$_{7-y}$ film, $\delta m_{ab}^{**}/m_{ab}^{**}$ was
found to be 5.5$\%$ upon replacing $^{16}$O with $^{18}$O (the $^{18}$O
concentration is about 95$\%$) \cite{Keller1}. If we assume that the oxygen-isotope
effect on $m_{ab}^{**}$ for HoBa$_{2}$Cu$_{4}$O$_{8}$ is similar to
that for the slightly underdoped
YBa$_{2}$Cu$_{3}$O$_{7-y}$ film, we expect that $\delta m_{ab}^{**}/m_{ab}^{**}$
= 4.3$\%$ in HoBa$_{2}$Cu$_{4}$O$_{8}$.

In the superconducting state or below $T^{*}$, the relaxation rate is
suppressed due to openning of the gap. Then the relaxation rate is
given by \cite{Temp}:

\begin{equation}
\Gamma \propto T\exp (-\Delta/2k_{B}T). \label{rate}
\end{equation}

The solid lines in Fig.~2 represent the best fits of Eq.~\ref{rate} to the data
below $T^{*}$.
The best fits yield $\Delta/k_{B}$ = 94.6$\pm$8.8~K for the $^{16}$O
sample and $\Delta/k_{B}$ = 286$\pm$19~K for the $^{18}$O
sample. Therefore, there is a giant oxygen-isotope effect on the
pseudogap, in agreement with Eq.~\ref{gapoie}. Substituting
$\delta\Delta$ = 16.5 meV and $\delta m_{ab}^{**}/m_{ab}^{**}$
= 4.3$\%$ into Eq.~\ref{gapoie}, we find $W$ = 0.256 eV and $t$ =
0.128 eV.
Using  $m_{ab}^{**}$ = $2\hbar^{2}/t a^{2}$, we calculate
$m_{ab}^{**}$ = 8.1 $m_{e}$, which is very close to that (8.3 $m_{e}$) inferred for
the slightly underdoped YBa$_{2}$Cu$_{3}$O$_{6.88}$ with $T_{c}$ =
87.9 K (see ref.~\cite{AlexPRB99}).

From the $\Delta$, $T_{c}$, and $W$ values of the $^{16}$O and $^{18}$O
samples, we can directly calculate $n_{p}(T_{c})$ to be 0.0117 and
0.0038 for the $^{16}$O and $^{18}$O
samples, respectively. Substituting $\delta T_{c}/T_{c}$ = $-$0.63$\%$
and $\delta m_{ab}^{**}/m_{ab}^{**}$
= 4.3$\%$ into Eq.~\ref{tcoie}, we obtain $\delta n_{p}(T_{c})/[x-n_{p}(T_{c})]$ =
$-$3.67$\%$.
With the $n_{p}(T_{c})$ values for the $^{16}$O and $^{18}$O
samples, we find $x$ = 0.227. Since the in-plane doping level $x$ in the
optimally-doped
YBa$_{2}$Cu$_{3}$O$_{6.95}$ was found to be 0.264 (ref.~\cite{Chma}), our inferred
doping level of 0.223 for HoBa$_{2}$Cu$_{4}$O$_{8}$ is consistent with
the fact that this compound is slightly underdoped. With $m_{ab}^{**}$ = 8.1 $m_{e}$
and $x$ = 0.227, we calculate $\lambda_{ab}(0)$ = 2248~\AA, which is close
to $\lambda_{a}(0)$ = 2000~\AA~for YBa$_{2}$Cu$_{4}$O$_{8}$
(ref.~\cite{Basov}).

The exponent $\alpha^{O}_{m^*}$ of the oxygen-isotope effect on $m_{ab}^{**}$
is calculated to be 0.46 for the slightly
underdoped YBa$_{2}$Cu$_{3}$O$_{7-y}$ and HoBa$_{2}$Cu$_{4}$O$_{8}$. Since the
exponent $\alpha^{Cu}$ of the copper-isotope effect on $T_{c}$ is
similar to  $\alpha^{O}$ in the underdoped regime \cite{Franck,ZhaoCu}, we expect that  $\alpha^{O}_{m^*}$
$\simeq$ $\alpha^{Cu}_{m^*}$ and $\alpha_{m^*}$ $\simeq$ 2$\alpha^{O}_{m^*}$. The
mass enhancement factor is then equal to $\exp (4\alpha^{O}_{m^*})$ =
6.3. The bare bandwidth $D$ = $2W\exp (4\alpha^{O}_{m^*})$ = 3.2 eV,
in quantitative agreement with the band-structure calculations \cite{Pick}.

In the absence of charge localization, as the case of the
stoichiometric HoBa$_{2}$Cu$_{4}$O$_{8}$, the Bose-Einstein
condensation temperature $T_{c}$ is given by \cite{alemot}

\begin{equation}
 k_{B}T_{c} = \frac{2tn_{b}(T_{c})}{1 + \ln(k_{B}T_{c}/2t_{c})},
 \label{tbe}
 \end{equation}
 where $t_{c}$ is related to the out-of-plane bipolaron mass
 $m_{c}^{**}$ as $t_{c}$ = $\hbar^{2}/2m_{c}^{**}d^{2}$. The out-of-plane bipolaron mass
 $m_{c}^{**}$ is deduced to be 518 $m_{e}$ using $m_{c}^{**}/m_{ab}^{**}$ = 64 at $T_{c}$
  (ref.~\cite{Huss}) and $m_{ab}^{**}$ = 8.1 $m_{e}$. Then  we calculate $t_{c}$ =
  0.160 meV. Substituting $t$ = 0.128 eV, $t_{c}$ = 0.160 meV, and $n_{b}(T_{c})$ = 0.1077
    into the above equation, we calculate $T_{c}$ = 78.9 K, in
    quantitative agreement with the measured $T_{c}$ (79.0 K). We
    further show that (see Supplementary Information \cite{sup}) Eq.~\ref{tbe} can also quantitatively explain the
    underdoped La$_{1.90}$Sr$_{0.10}$CuO$_{4}$ with $T_{c}$ = 29~K but overestimates the
    $T_{c}$'s of some optimally doped samples. This implies that optimally doped cuprates are in the crossover regime from bipolaronic real-space pairing  to the
    Cooper pairing of polarons \cite{alebook}.

Apart from the striking isotope effects explained quantitatively here there is
abundant independent evidence in favor of bipolarons and the
Bose-Einstein condensation  in underdoped cuprate superconductors. In
particular, the parameter-free estimates of the Fermi energy using
the magnetic-field penetration depth \cite{nonadi} and the magnetic
quantum oscillations \cite{osc} yielded a very low value (below 50
meV) supporting the real-space pairing in underdoped cuprate
superconductors. Magnetotransport and thermal magnetotransport data
strongly support preformed bosons in cuprates. In particular, many
high-magnetic-field studies revealed the non-BCS upward curvature
of the upper critical field $H_{c2}(T)$ as a function of temperature
\cite{hc2} as predicted for the Bose-Einstein condensation of
charged bosons in the magnetic field \cite{hc2asa}. The Lorenz
number differs significantly from the conventional Sommerfeld value
of the standard Fermi-liquid theory because the carriers are double charged
bosons \cite{lorASA}.  Direct measurements of the Lorenz number
using the thermal Hall effect just
above $T_{c}$ \cite{lor} produce its value, which is about the same as that
predicted by the bipolaron model. The
unusual normal-state diamagnetism uncovered by torque magnetometery
has been convincingly explained as the normal state (Landau)
diamagnetism of charged bosons \cite{dia}. Single polarons,
localised within an impurity band-tail, coexist with bipolarons in
the charge-transfer doped Mott-Hubbard insulator, where the chemical
potential is pinned within the charge-transfer gap due to bipolaron
formation. This band-tail model accounts for two energy scales in
ARPES and in the extrinsic and intrinsic tunnelling, their
temperature and doping dependence, and for the asymmetry and
inhomogeneity of extrinsic tunnelling spectra of cuprates
\cite{tunarpes}.

\section*{Supplementary Information$-$Isotope effects in  high-T$_c$
cuprate superconductors: Ultimate proof for  bipolaron theory of
superconductivity}

\noindent{\bf 1 Microscopic derivation of the low-energy
band-structure}

 In highly polarizable ionic lattices like
cuprate superconductors both the Coulomb repulsion and the
Fr\"ohlich electron-phonon interaction (EPI) are quite strong (of
the order of 1 eV) compared with the low Fermi energy of doped
carriers because of a poor screening by non- or near-adiabatic
carriers \cite{alebra}. In those conditions the BCS-Eliashberg
theory \cite{me} breaks down because of the polaronic collapse of
the electron bandwidth \cite{ale0} so that one has to apply a
non-adiabatic small polaron theory \cite{alecond}.

Here we sketch the microscopic derivation of the low-energy band
structure, Fig.~1, using the analytical multi-polaron theory in the
strong-coupling regime for highly polarisable lattices (more details
are found in Refs.~\cite{alecond}).

 Quantitative calculations of the
interaction matrix elements can be performed from pseudopotentials
using the density functional theory (DFT) \cite{bauer}. On the other
hand, one can express the bare Coulomb repulsion and EPI through
material parameters rather than computing them from first principles
in many physically important cases \cite{mahan}. In
particular, for a polar coupling to longitudinal optical phonons (the Fr\"{o}%
hlich EPI), which is the major EPI in polar crystals, both the
momentum dependence of the matrix element, $M(\mathbf{q})$, and its
magnitude are well known, $|M(\mathbf{q})|=\gamma (q))\hbar \omega
_{0}/\sqrt{2N}$ with a dimensionless $\gamma (q)=\sqrt{4\pi
e^{2}/\kappa \Omega \hbar \omega _{0}q^{2}}$, where $\Omega $ is a
unit cell volume, $N$ is the number of
unit cells in a crystal, $\omega _{0}$ is the optical phonon frequency, and $%
\kappa =\epsilon _{\infty }\epsilon _{0}/(\epsilon _{0}-\epsilon _{\infty })$%
. The high-frequency, $\epsilon _{\infty }$ and the static,
$\epsilon _{0}$ dielectric constants are both measurable in a parent
polar insulator.

The
dielectric response function of strongly correlated electrons is \emph{%
apriori} unknown. Hence one has to start with a generic Hamiltonian
including \emph{unscreened} Coulomb and Fr\"ohlich interactions
operating on the same scale since any ad-hoc assumption on their
range and relative magnitude might fail,
\begin{eqnarray}
H &=-&\sum_{i,j} (T_{ij}\delta_{ss^\prime}+\mu \delta _{ij})
c_{i}^{\dagger
}c_{j} +{\frac{1}{{2}}}\sum_{i\neq j}{\frac{e^2}{{\epsilon_\infty |\mathbf{%
m-n}|}}}\hat{n}_{i}\hat{n}_j+\cr && \sum_{\mathbf{q},i}\hbar \omega _{0}\hat{%
n}_{i}\left[ u(\mathbf{m,q} ) d_{\mathbf{q} }+H.c.\right]+H_{ph}.
\label{hamiltonian}
\end{eqnarray}
Here $T_{ij}\equiv T(\mathbf{m-n})$ is the bare hopping integral,
$\mu$ is the chemical potential, $i=\mathbf{m},s$ and
$j=\mathbf{n},s^{\prime }$
include both site $(\mathbf{m,n})$ and spin $(s,s^{\prime })$ states, $u(%
\mathbf{m,q})= (2N)^{-1/2}\gamma(q)\exp(i \mathbf{q \cdot m})$, $c_{i}, d_{%
\mathbf{q} }$ are electron and phonon operators, respectively, $\hat{n}%
_{i}=c^\dagger_i c_i$ is a site occupation operator, and $H_{ph}=\sum_{%
\mathbf{q}}\hbar \omega _{0}(d_{\mathbf{q} }^{\dagger
}d_{\mathbf{q}}+1/2)$ is the polar vibration energy.

In highly polarisable lattices with
$\epsilon_{0}~\rightarrow~\infty$ the familiar Lang-Firsov (LF)
\cite{fir} canonical transformation $e^{S}$ is
particulary instrumental with $S=-\sum_{\mathbf{q},i}\hat{n}_{i}\left[ u(%
\mathbf{m,q})d_{\mathbf{q}}-H.c.\right] $. It shifts the ions to new
equilibrium positions changing the phonon vacuum, and removes most of \emph{%
both }interactions from the transformed Hamiltonian, $\tilde{H}=e^{S}He^{-S}$%
,
\begin{equation}
\tilde{H}=-\sum_{i,j}(\hat{\sigma}_{ij}\delta _{ss^{\prime }}+\tilde{\mu}%
\delta _{ij})c_{i}^{\dagger }c_{j}+H_{ph},  \label{trans}
\end{equation}%
where $\hat{\sigma}_{ij}=T({\mathbf{m-n}})\hat{X}_{i}^{\dagger
}\hat{X}_{j}$ is the renormalised hopping integral involving the
multi-phonon transitions
described with $\hat{X}_{i}=\exp \left[ \sum_{\mathbf{q}}u(\mathbf{m,q})d_{%
\mathbf{q}}-H.c.\right] $, and $\tilde{\mu}=\mu +E_{p}$ is the
chemical potential shifted by the polaron level shift,
\begin{equation}
E_{p}={\frac{2\pi e^{2}}{{\kappa }}}\int_{BZ}{\frac{d^{3}q}{{(2\pi )^{3}q^{2}%
}}}.  \label{shift}
\end{equation}%
Here, the integration goes over the Brillouin zone (BZ) and
$E_{p}=0.647$ eV in La$_{2}$CuO$_{4}$ \cite{alebra}. The
electron-phonon coupling constant is defined as $\lambda
=2E_{p}N(0)$. In the case of 2D carriers with a
constant bare density of states, $N(0)=ma^{2}/2\pi \hbar ^{2}$ per spin, Eq.(%
\ref{shift}) places cuprates in the intermediate to strong-coupling
regime, $\lambda \gtrsim 0.5$, if the bare band mass $m>m_{e}$ (here
$a$ is the in-plane lattice constant).

The number of virtual phonons in the polaron cloud is large in
oxides and some other polar lattices, $E_p/\hbar \omega_0 > 1$ with
the characteristic (oxygen) optical phonon frequency $\hbar \omega_0
\lesssim 80$ meV, so that multi-phonon vertexes are essential in the
expansion of the hopping operator $\hat{\sigma}_{ij}$. To deal with
this challenging problem
let us single out the coherent hopping in Eq.(\ref{trans}) averaging $\hat{%
\sigma}_{ij}$ with respect to the phonon vacuum, and consider the
remaining terms as perturbation, $\tilde{H}=H_{0}+H_{p-ph}$. Here
\begin{equation}
H_{0}=-\sum_{i,j}(t_{ij}\delta_{ss^\prime}+\tilde{\mu}
\delta_{ij})c_{i}^{\dagger }c_{j}+H_{ph}
\end{equation}
 describes free phonons and polarons coherently propagating in
a narrow band with the exponentially diminished hopping integral, $t_{ij}=T(%
\mathbf{m}-\mathbf{n})\exp[-g^2(\mathbf{m}-\mathbf{n})]$,
\begin{equation}
g^2(\mathbf{m})={\frac{1}{{2N}}}\sum_\mathbf{q}\gamma(q)^2
[1-\cos(\mathbf{q }\cdot \mathbf{m})],  \label{t}
\end{equation}
and
\begin{equation}
H_{p-ph}=\sum_{i,j}(t_{ij}-\hat{\sigma}_{ij})\delta_{ss^\prime}c_{i}^{%
\dagger }c_{j}
\end{equation}
 is the residual polaron-multiphonon interaction, which is a
perturbation at large $\lambda$. In the diagrammatic technique the
corresponding vertexes have any number of phonon lines. The
second-order in $H_{p-ph}$ polaron self-energy ($\Sigma_p \approx -
E_p/2z \lambda^2$) and the phonon self-energy ($\Sigma_{ph}\approx
-x\hbar\omega_0/z \lambda^2$)  are small, if $\lambda \gg
1/\sqrt{2z} $ \cite{ale92}(here $z$ is the lattice coordination
number and $x$ is the atomic density of carriers). Hence the
perturbation expansion in $1/\lambda $ is applied. Importantly there
is no structural instability in the strong coupling regime since
$|\Sigma_{ph}|\ll \hbar \omega_0$ \cite{ale92}.

The LF transformation, Eq.~(\ref{trans}) is exact for any adiabatic ratio $%
\hbar \omega_0/T(a)$. However, if the perturbation expansion in
$1/\lambda$ is restricted by lowest orders, then it significantly
overestimates polaron masses in the adiabatic regime, $\hbar
\omega_0/T(a) < 1$, for the case of the short-range (Holstein) EPI
 (here $T(a)$ is the nearest-neighbor bare hopping
integral). The polaronic band narrowing factor, $\exp(-g^2)$ becomes
very small for this EPI in the strong-coupling regime, which would
eliminate any possibility of high temperature superconductivity and
even metallicity of the small Hosltein polarons.

However in the case of the long-range (Fr\"ohlich) EPI, Quantum
Monte-Carlo simulations \cite{alekor} show that the LF
transformation provides numerically accurate polaron masses already
in the zero order of the inverse-coupling expansion both in the
adiabatic regime as well as in the non-adiabatic one for \emph{any }
strength of the Fr\"ohlich EPI. Moreover, such small  polarons
\cite{alekor} and  small bipolarons \cite{ale96} are perfectly
mobile in the relevant range of the coupling and the adiabatic
ratio.

The perturbation $H_{p-ph}$ has no diagonal matrix elements with
respect to phonon occupation numbers. Hence it can be removed from
the Hamiltonian in the first order using a second canonical
transformation ${\mathcal{H}} =e^{S_2}\tilde{H} e^{-S_2}$ with
$(S_2)_{n^\prime n}=\sum_{i,j} \langle n^\prime
|(\hat{\sigma}_{ij}-t_{ij})c_{i}^{\dagger
}c_{j}|n\rangle/(E_{n^\prime}-E_n)$, where $E_n, E_{n^\prime}$ and $%
|n\rangle, |n^\prime\rangle$ are the energy levels and the eigenstates of $%
H_0$, respectively. Taking into account that the polaron Fermi
energy is small compared with the phonon energy at strong coupling
and/or sufficiently
low doping \cite{alebra}, one can neglect the polaron contribution to $%
E_{n^\prime}-E_n \approx \hbar \omega_0 \sum_{\mathbf{q}} n^\prime _{\mathbf{%
q}}$ and project the second-order in $1/\lambda$ Hamiltonian
$\mathcal{H}$ onto the phonon vacuum $|0\rangle$ with the following
result
\begin{eqnarray}
\mathcal{H}&=&-\sum_{i,j}(t_{ij}\delta_{ss^\prime}+\tilde{\mu}
\delta_{ij})c_{i}^{\dagger }c_{j}-\cr &&
\sum_{\mathbf{mn}\mathbf{m}^\prime
\mathbf{n}^\prime,ss^\prime} V_{\mathbf{m n}}^{\mathbf{m}^\prime \mathbf{n}%
^\prime}c_{\mathbf{m}s}^{\dagger}c_{\mathbf{n}s}c_{\mathbf{m}^\prime
s^\prime}^{\dagger}c_{\mathbf{n}^\prime s^\prime},
\end{eqnarray}
where
\begin{equation}
V_{\mathbf{mn}}^{\mathbf{m}^\prime \mathbf{n}^\prime}=iT_{ij}T_{i^%
\prime j^\prime}\int_0^{\infty}dt e^{-\delta t}\langle 0|[\hat{X}%
^{\dagger}_i(t)\hat{X}_j(t)-1]\hat{X}^{\dagger}_{i^\prime} \hat{X}%
_{j^\prime}|0\rangle, \label{int}
\end{equation}
 and $\hat{X}^{\dagger}_i(t)$ is the Heisenberg
multi-phonon operator obtained by replacing $d_{q}$ in
$\hat{X}^{\dagger}_i$ with $d_{q}\exp (i \omega_0 t)$. Calculating
the integral, Eq.(\ref{int})
 with $\delta
\rightarrow +0$ yields
\begin{equation}
V_{\mathbf{mn}}^{\mathbf{m}^\prime \mathbf{n}^\prime}={\frac{%
t_{ij}t_{i^\prime j^\prime}}{{\hbar \omega_0}}} \sum_{k=1}^\infty {\frac{f(%
\mathbf{mn},\mathbf{m}^\prime \mathbf{n}^\prime)^k}{{k! k}}},
\label{V2}
\end{equation}
where $f(\mathbf{mn},\mathbf{m}^\prime\mathbf{n}^\prime)=(1/2N) \sum_\mathbf{%
q}\gamma(q)^2 [\cos(\mathbf{q} \cdot (\mathbf{m}-\mathbf{n}^\prime))+\cos(%
\mathbf{q} \cdot (\mathbf{n}-\mathbf{m}^\prime))-\cos(\mathbf{q} \cdot (%
\mathbf{m}-\mathbf{m}^\prime))-\cos(\mathbf{q} \cdot (\mathbf{n}-\mathbf{n}%
^\prime))]$.

All matrix elements, Eq.~(\ref{V2}), of the polaron-polaron
interaction are small compared with the polaron kinetic energy
except the \emph{exchange} interaction, $J_p(\mathbf{m}-\mathbf{n})
\equiv V_{\mathbf{mn}}^{\mathbf{n}
\mathbf{m}}$ such that $f(\mathbf{mn},\mathbf{m}^\prime\mathbf{n}%
^\prime)=2g^2(\mathbf{m}-\mathbf{n})$. Using $\sum_{k=1}^\infty
y^k/k! k= -C-\ln(y)+Ei^\star(y)$ with $C\approx 0.577$ and
$Ei^\star(y) \approx e^y/y$
(for large $y$) one obtains a substantial $J_p(\mathbf{m})=T^2(\mathbf{m}%
)/2g^2(\mathbf{m}) \hbar \omega_0 $, which is  larger than the
nearest-neighbour  polaron hopping integral, $t(a)/J_p \propto
2\hbar \omega_0 g^2 e^{-g^2}/T(a) < 1$.  Keeping only this exchange
we finally arrive with the polaronic "t-J$_p$" Hamiltonian
\cite{alecond},
\begin{eqnarray}
\mathcal{H}&=&-\sum_{i,j}(t_{ij}\delta_{ss^\prime}+\tilde{\tilde{\mu}}
\delta_{ij})c_{i}^{\dagger }c_{j} \cr &+&2 \sum_{\mathbf{m} \neq
\mathbf{n}}
J_p(\mathbf{m}-\mathbf{n}) \left(\vec{S}_\mathbf{m} \cdot \vec{S}_\mathbf{n}+%
{\frac{1}{{4}}}\hat{n}_\mathbf{m}\hat{n}_\mathbf{n}\right),
\label{tJ}
\end{eqnarray}
where $\vec{S}_\mathbf{m}=(1/2)\sum_{s,s^\prime}c^\dagger_{\mathbf{m}s}\vec{%
\tau}_{ss^\prime} c_{\mathbf{m}s^\prime}$ is the spin 1/2 operator ($\vec{%
\tau}$ are the Pauli matrices), $\hat{n}_\mathbf{m}=\sum_s \hat{n}_i$, and $%
\tilde{\tilde{\mu}}=\tilde{\mu}+\sum_{\mathbf{m}}J_p(\mathbf{m})$ is
the chemical potential further renormalized by $H_{p-ph}$.

There is a striking difference between this polaronic t-J$_p$
Hamiltonian and the familiar t-J model derived from the repulsive
Hubbard U Hamiltonian in the limit $U\gg t$ omitting the so-called
three-site hoppings and EPI \cite{tJ}. The latter model acts in a
projected Hilbert space constrained to no double occupancy.  On the
contrary in the polaronic t-J$_p$ Hamiltonian, Eq.~(\ref{tJ}) there
is no constraint on the double on-site occupancy since the Coulomb
repulsion is negated by the Fr\"ohlich EPI. The polaronic hopping
integral $t(a)$ leads to the coherent (bi)polaron band and the
antiferromagnetic exchange of purely phononic origin $J_p$  bounds
polarons into small superlight inter-site bipolarons. Last but not
least the difference is in the "+" sign in
the last term of Eq.~(\ref{tJ}) proportional to $\hat{n}_\mathbf{m}\hat{n}_%
\mathbf{n}$, which protects the ground superconducting state from
the bipolaron clustering, in contrast with the "-" sign in the
similar term of the standard t-J model, where the phase separation
is expected at sufficiently large J \cite{kiv}.

The polaronic t-J$_p$ Hamiltonian, Eq.~(\ref{tJ}) is analytically
solvable in the limit of sufficiently low atomic density of carriers
\cite{alecond}. Neglecting the first term in $\mathcal{H}$, which is
the polaron kinetic energy proportional to $t(a) < J_p$, one can
readily diagonalise the remaining spin-exchange part of the
Hamiltonian. Its ground state is an
ensemble of inter-site singlet bipolarons with the binding energy $%
\Delta_b=J_p$ localised on nearest neighbor sites. Such small
bipolarons
repel each other and single polarons via a short-range repulsion of about $%
J_p$.

The kinetic energy operator in Eq.~(\ref{tJ}) connects singlet
configurations in the first and higher orders with respect to the
polaronic hopping integrals. Taking into account only the
lowest-energy degenerate singlet configurations and discarding all
other configurations one can project the t-J$_p$ Hamiltonian onto
the inter-site bipolaronic Hamiltonian
using the bipolaron annihilation operators $B_{\mathbf{m}}=2^{-1/2} (c_{%
\mathbf{m}\uparrow}c_{\mathbf{m}+\mathbf{a}\downarrow}-c_{\mathbf{m}%
\downarrow}c_{\mathbf{m}+\mathbf{a}\uparrow})$, where $\mathbf{a}$
connects nearest neighbors \cite{ale96}. Such inter-site bipolarons
are perfectly mobile since they tunnel via single-polaron
transitions \cite{ale96,hague}. At finite temperatures single
polarons, thermally excited above the pseudogap, coexists with these
bipolarons as shown in Fig.~1 of the main text.

Small bipolarons are hard-core bosons with the short-range repulsion
 and a huge anisotropy of their effective mass
since their inter-plane hopping is possible only in the second order
of the polaron hopping integral \cite{alekabmot}. The occurrence of
superconductivity in bipolaronic  systems   is not controlled by the
pairing strength, but by the phase coherence among the electron
pairs below the Bose-Einstein condensation temperature
\cite{aleran}.

\noindent{\bf 2 Parameter-free calculations of the Bose-Einstein
consensation temperatures of some cuprate superconductors}

In the absence of charge localization, the Bose-Einstein
condensation temperature $T_{c}$ is given by \cite{alemot}

\begin{equation}
 k_{B}T_{c} = \frac{2tn_{b}(T_{c})}{1 + \ln(k_{B}T_{c}/2t_{c})},
 \label{tbe}
 \end{equation}
 where $t$ is the bipolaron half-bandwidth and $t_{c}$ is related to the out-of-plane bipolaron mass
 $m_{c}^{**}$ as $t_{c}$ = $\hbar^{2}/2m_{c}^{**}d^{2}$ ($d$ is the inter-plane distance). The above
 equation can be written in terms of the measurable parameters such as
 the inplane
 penetration depth and the supercarrier mass anisotropy constant
 $\gamma^2$ = $m_{c}^{**}/m_{ab}^{**}$,
 \begin{eqnarray}
 k_{B}T_{c} &=& \frac{dh^{2}c^{2}}{16\pi^{3}e^{2}\lambda_{ab}^{2}(0)}\times \nonumber \\
 && [1 +
 \ln
 (\frac{32\pi^{3}xe^{2}k_{B}T_{c}\lambda_{ab}^{2}(0)\gamma^{2}d}{a^{2}h^{2}c^{2}})]^{-1}.
 \label{tbe1}
 \end{eqnarray}
It is worth noting that the $T_{c}$ value calculated from
 Eq.~\ref{tbe1} should be slightly overestimated due to the fact that
 $n_{b}(T_{c})$ is slightly lower than $n_{b}(0)$ = $x/2$.  For
 La$_{1.90}$Sr$_{0.10}$CuO$_{4+y}$, $\lambda_{ab}(0)$ = 291 nm
 (ref.~\cite{Uemura}),
 $\gamma$ = 43 (ref.~\cite{Will}), and $x$ = 0.1. These parameters lead to $T_{c}$ =
 31.8~K, in quantitative agreement with the measured value of 29 K. For
 the optimally doped YBa$_{2}$Cu$_{3}$O$_{6.95}$ with $T_{c}$ = 93 K, $\lambda_{ab}(0)$
 = 1600~\AA~(ref.~\cite{Basov}), $x$ = 0.264 (ref.~\cite{Chma}), and
 $\gamma$ = 8 (ref.~\cite{Taj}), so $T_{c}$ is calculated to
 be 162~K, which is  higher than the measured value of 93 K pointing to the BEC-BCS crossover \cite{ale0}.


\begin{thebibliography}{99}

\bibitem{bedmul} Bednorz, J. G., and M\"uller, K. A. Possible high
T$_c$ superconductivity in the Ba-La-Cu-O system. Z. Phys. B
\textbf{64}, 189 (1986).

\bibitem{kam} Kammerling-Onnes, H. Further experiments with liquid helium. D. On the change of electric resistance of pure metals at very low temperatures, etc. V. The disappearance of the resistance of mercury.
\emph{ Comm. Phys. Lab. Univ. Leiden}, No. 122b (1911).

\bibitem{max} Maxwell, E. Isotope Effect in the Superconductivity of Mercury. \emph{Phys. Rev.} \textbf{78}, 477
(1950).
\bibitem{rey} Reynolds, C. A.,  Serin, B.,   Wright, W. H.,  and
Nesbitt, L. B. Superconductivity of Isotopes of Mercury. \emph{Phys
. Rev.} \textbf{78}, 487 (1950).

\bibitem{fro}Fr\"ohlich, H. Theory of the Superconducting State. I.
The Ground State at the Absolute Zero of Temperature. \emph{Phys.
Rev.} \textbf{79}, 845 (1950).

\bibitem{coop}  Cooper, L. N. Bound Electron Pairs in a Degenerate Fermi
Gas. \emph{Phys. Rev.} \textbf{104}, 1189 (1950).

\bibitem{bar} Bardeen, J., Cooper, L. N., and Schrieffer, J. R. Theory
of superconductivity. Phys. Rev. \textbf{108 }, 1175204 (1957).


\bibitem{polarons} Zhao, G. M. Polarons in Colossal Magnetoresistive
and  High-Temperature  Superconducting Materials. \emph{Polarons in
Advanced Materials}, ed. Alexandrov A S (Dordrecht: Springer, 2007)
pp 569-597; Bussmann-Holder, A. and Keller, H. Polaron Effects in
High-Temperature Cuprate Superconductors. \emph{ibid} pp 599-621.

\bibitem{ZhaoYBCO95}Zhao, G. M. and Morris, D. E. Observation of possible
oxygen isotope effect on the effective mass of carriers in
YBa$_{2}$Cu$_{3}$O$_{6.94}$. \emph{Phys. Rev. B} {\bf 51},
R16487-16490 (1995).

\bibitem{ZhaoLSCO95}Zhao, G. M.,  Singh, K. K.,  Sinha, A. P. B., and
 Morris, D. E. Oxygen isotope effect on the effective mass of carriers
from magnetic measurements on La$_{2-x}$Sr$_{x}$CuO$_{4}$.\emph{
Phys. Rev. B} \textbf{52}, 6840-6844 (1995).

\bibitem{ZhaoNature97}Zhao, G. M.,  Hunt, M. B.,  Keller, H., and
M\"uller, K. A. Evidence for polaronic supercarriers in the copper
oxide superconductors La$_{2-x}$Sr$_{x}$CuO$_{4}$. \emph{Nature
(London)} {\bf 385}, 236-239 (1997).

\bibitem{ZhaoJPCM98} Zhao, G. M.,  Conder, K.,  Keller, H., and
M\"uller, K. A. Oxygen isotope effects in
La$_{2-x}$Sr$_{x}$CuO$_{4}$: evidence for polaronic charge carriers
and their condensation. \emph{J. Phys.: Condens. Matter} {\bf 10},
9055-9066 (1998).

\bibitem{Zhao01} Zhao, G.  M.,   Kirtikar, V. and
Morris, D.  E. Isotope effects and possible pairing mechanism in
optimally doped cuprate superconductors.  \emph{Phys.  Rev.  B} {\bf
63}, R220506-220509 (2001).

\bibitem{Keller1}  Khasanov, R. {\em et al.}.  Direct Observation of the Oxygen
Isotope Effect on the In-Plane Magnetic Field Penetration Depth in
Optimally Doped YBa$_{2}$Cu$_{3}$O$_{7}$.  \emph{Phys.  Rev.  Lett.}
{\bf 92}, 057602-057605 (2004).




\bibitem{lanzara} Lanzara, A. \emph{et al.}  Evidence for ubiquitous strong
electron-phonon coupling in high-temperature superconductors.
\emph{Nature (London)} 412 510 (2001); Meevasana, et al.  Doping
Dependence of the Coupling of Electrons to Bosonic Modes in the
Single-Layer High-Temperature Bi$_2$Sr$_2$CuO$_6$ Superconductor.
\emph{Phys. Rev. Lett.} \textbf{96}, 157003 (2006).


\bibitem{mic} Mihailovic, D.,   Foster, C. M.,  Voss, K., and Heeger, A. J. Application of the polaron-transport theory to $\sigma(\omega)$ in Tl$_2$Ba$_2$Ca$_{1-x}$Gd$_x$Cu$_2$O$_8$, YBa$_2$Cu$_3$O$_{7-\delta}$, and
La$_{2-x}$Sr$_x$CuO$_4$. \emph{Phys. Rev. B} \textbf{42}, 7989
(1990)

\bibitem{ega}  Sendyka, T. R.,  Dmowski, W.,  Egami, T.,  Seiji, N.,  Yamauchi, H., and
Tanaka, S. Temperature dependence of the local structure
of YBa$_{2}$Cu$_{4}$O$_{8}$. \emph{Phys. Rev. B} \textbf{51}, 6747 (1995).

\bibitem{rez}  Reznik, D.,   Pintschovius, L.,   Ito, M.,  Iikubo, S., Sato, M.,
Goka, H.,   Fujita, M.,   Yamada, K.,   Gu, G. D., and Tranquada, J.
M.,  Electron-phonon coupling reflecting dynamic charge
inhomogeneity in copper oxide superconductors. \emph{Nature
(London)} \textbf{440}, 1170 (2006).

\bibitem{Zhao07} Zhao, G. M. Strong coupling to multiple phonon modes in high-temperature
superconductors. \emph{Phys. Rev. B} {\bf 75}, 214507-214510 (2007).

\bibitem{Boz08}Shim. H.,
Chaudhari, P., Logvenov, G. and  Bozovic, I. Electron-Phonon
Interactions in Superconducting La$_{1.84}$Sr$_{0.16}$CuO$_{4}$
Films. \emph{Phys. Rev. Lett.} {\bf 101}, 247004-247007 (2008).
\bibitem{ZhaoPRL09} Zhao, G. M. Fine structure in the tunneling spectra of electron-doped
cuprates: No coupling to the magnetic resonance mode. \emph{Phys.
Rev. Lett.} {\bf 103}, 236403-236406 (2009).

\bibitem{dev} Tempere, J., and  Devreese, J. T. Optical absorption of an interacting many-polaron
gas.
\emph{Phys. Rev. B} \textbf{64}, 104504 (2001).

\bibitem{asa}  Alexandrov, A. S. \emph{Phys. Rev. B} \textbf{53}, 2863 (1996).
\bibitem{muller:2000} M\"uller, K. A. On the superconductivity in hole doped
cuprates. \emph{J. Phys.: Condens. Matter} \textbf{19}, 251002
(2007).
\bibitem{alebook}  Alexandrov, A. S. \emph{Theory of Superconductivity: From Weak to Strong Coupling}
(IoP Publishing, Bristol,  2003).
\bibitem{sup} See supplementary Information at http://


\bibitem{Muller2011}Weyeneth, S. and  M\"uller, K. A. Oxygen Isotope Effect in
Cuprates Results from Polaron-induced Superconductivity. \emph{J.
Supercond. Nov. Magn.}  {\bf 24}, 1235-1239 (2011).

\bibitem{Keller2} Khasanov, R.,  Str\''assle, S.,  Conder, K.,  Pomjakushina,
 E.,  Bussmann-Holder, A.,
and  Keller, H. Universal correlations of isotope effects in
Y$_{1-x}$Pr$_{x}$Ba$_{2}$Cu$_{3}$O$_{7-\delta}$.  \emph{Phys. Rev.
B} {\bf 77}, 104530-104534 (2008).

\bibitem{Tallon} Tallon, J. L.,  Islam, R. S.,  Storey, J.,   Williams, G.V.
M., and Cooper, J. R. Isotope Effect in the Superfluid Density of
High-Temperature Superconducting Cuprates: Stripes, Pseudogap, and
Impurities. \emph{Phys. Rev. Lett.} {\bf 94}, 237002-237005 (2005).

\bibitem{aleiso} Alexandrov, A. S. Transition from Fermi liquid to
charged Bose liquid: A possible explanation of the isotope shift in
high-Tc oxides. \emph{Phys. Rev. B} {\bf 46}, 14932 (1992).



\bibitem{Temp} Rubio Temprano, D.,  Mesot, J.,  Janssen, S.,  Conder,
K.,  Furrer, A., Mutka, H., and
M\"uller, K. A. Large Isotope Effect on the Pseudogap in the
High-Temperature Superconductor HoBa$_{2}$Cu$_{4}$O$_{8}$.
\emph{Phys. Rev. Lett.} {\bf 84}, 1990-1993 (2000).

\bibitem{hague}  Hague, J. P.,  Kornilovitch, P. E.,   Samson, J. H. and
Alexandrov, A. S. Superlight Small Bipolarons in the Presence of a
Strong Coulomb Repulsion.  \emph{Phys. Rev. Lett.} \textbf{98},
037002-037005 (2007)




\bibitem{AlexPRB99} Alexandrov, A. S. and  Kabanov, V. V. Parameter-free expression
for superconducting $T_{c}$ in cuprates. \emph{Phys. Rev. B} {\bf
59}, 13628-13631 (1999).




\bibitem{Chma}Chmaissem, O., Eckstein,  Y., and  Kuper, C. G. Structure
and a bond-valence-sum study of the 1-2-3 superconductors
(Ca$_{x}$La$_{1-x}$)(Ba$_{1.75-x}$La$_{0.25+x}$)Cu$_{3}$O$_{y}$ and
YBa$_{2}$Cu$_{3}$O$_{y}$. \emph{Phys. Rev. B} {\bf 63},
174510-174519 (2001).

\bibitem{Basov}Basov, D. N. {\em et al.}, In-Plane Anisotropy
of the Penetration Depth in YBa$_{2}$Cu$_{3}$O$_{7-x}$ and
YBa$_{2}$Cu$_{4}$O$_{8}$ Superconductors. \emph{Phys.  Rev.  Lett.}
{\bf 92}, 598-601 (1995).

\bibitem{Huss}Hussey, N. E.,  Nozawa, K.,  Takagi, H.,  Adachi, S.,  and  Tanabe, K.
Anisotropic resistivity of YBa$_{2}$Cu$_{4}$O$_{8}$: Incoherent-to-metallic crossover in the out-of-plane transport.
\emph{Phys. Rev. B} {\bf 56}, R11423-11426 (1997).

\bibitem{Franck} Franck, J. P.,  Harker, S., and  Brewer, J. H. Copper and oxygen isotope effects in
La$_{2-x}$Sr$_{x}$CuO$_{4}$. Phys. Rev. Lett. {\bf 71}, 283-286 (1993).

\bibitem{ZhaoCu}Zhao, G. M., Kirtikar, V., Singh, K. K., Sinha, A. P.
B., Morris, D. E., and Inyushkin, A. V.  Large copper isotope effect in oxygen depleted YBa$_{2}$Cu$_{3}$O$_{7-x}$: Importance of Cu-dominated
phonon modes in the pairing mechanism. \emph{Phys. Rev. B} {\bf 54},
14956-14959 (1996).

\bibitem{Pick}Pickett, W. E. Electronic structure of the high-temperature
oxide superconductors. \emph{Rev. Mod. Phys.} {\bf 61}, 433-512 (1989).

\bibitem{alemot}Alexandrov, A. S. and  Mott, N. F. {\em Polarons and Bipolarons}, p.~144
 (World Scientific, Singapore, 1995).


\bibitem{nonadi}Alexandrov, A. S.  Nonadiabatic polaronic superconductivity in
MgB$_{2}$ and cuprates. \emph{Physica C} \textbf{363}, 231 (2001).

\bibitem{osc} Doiron-Leyraud, N. \emph{et al.}  Quantum oscillations and the Fermi
surface in an underdoped high-$T_{c}$ superconductor.\emph{Nature
(London)} \textbf{44}, 7565 (2007).

\bibitem{hc2}Zavaritsky, V. N., Kabanov, V. V., and Alexandrov, A.
S.  Universal upper critical field of unconventional
superconductors. \emph{Europhys. Lett.} \textbf{60} 127 (1998).

\bibitem{hc2asa}Alexandrov, A. S. Bose-Einstein condensation of
charged bosons in a magnetic field. \emph{Phys. Rev. B} \textbf{48},
10571 (1993).

\bibitem{lorASA} Alexandrov, A. S., and Mott, N. F.  Thermal transport
in a charged Bose gas and in high-Tc oxides.  \emph{Phys. Rev.
Lett.} \textbf{71} 1075-1078 (1993).

\bibitem{lor}Zhang, Y., Ong, N. P., Xu, Z. A., Krishana, K., Gagnon, R., and Taillefer,
L. Determining the Wiedemann-Franz Ratio from the Thermal Hall
Conductivity: Application to Cu and YBa$_2$Cu$_3$O$_{6.95}$.
\emph{Phys. Rev. Lett.} {\bf 84} 2219-2222 (2000).

\bibitem{dia} Alexandrov, A. S.  Normal-State Diamagnetism of
Charged Bosons in Cuprate Superconductors. \emph{Phys. Rev. Lett.}
\textbf{96 }, 147003-147006 (2006).

\bibitem{tunarpes} Alexandrov, A. S. and Beanland, J.  Superconducting
Gap, Normal State Pseudogap, and Tunneling Spectra of Bosonic and
Cuprate Superconductors. \emph{Phys. Rev. Lett.} \textbf{104}
026401-026403
(2010).


\end{thebibliography}

\begin{thebibliography}{99}


\bibitem{alebra}  Alexandrov A. S. and  Bratkovsky A. M., Phys. Rev. Lett.
\textbf{105}, 226408 (2010).

\bibitem{me}  Eliashberg G. M., Zh. Eksp. Teor. Fiz. \textbf{39}, 1437 (1960)
[Sov. Phys.--JETP  \textbf{12}, 1000 (1960)].

\bibitem{ale0}  Alexandrov A. S., Zh. Fiz. Khim. \textbf{57}, 273 (1983)
[Russ. J. Phys. Chem. \textbf{57}, 167 (1983)].



\bibitem{alecond} A. S. Alexandrov, Theory of High Temperature Superconductivity in Doped Polar
Insulators, ArXiv/0227756.




\bibitem{bauer}  Bauer T. and  Falter C., Phys. Rev. B \textbf{80}, 094525
(2009).

\bibitem{mahan} Mahan  G. D., \textit{Many-Particle Physics} (Plenum, New
York 1990).

\bibitem{fir}  Lang I.~G. and  Firsov Y.~A., Zh. Eksp. Teor. Fiz. \textbf{43},
1843 (1962) [Sov. Phys. JETP \textbf{16}, 1301 (1962)].


\bibitem{ale92}  Alexandrov A. S., Phys. Rev. B \textbf{46}, 2838 (1992).

\bibitem{alekor}  Alexandrov A. S. and  Kornilovitch P. E., Phys. Rev. Lett.
\textbf{82}, 807 (1999).
\bibitem{ale96}  Alexandrov A. S.,  Phys. Rev. B \textbf{53}, 2863 (1996).

\bibitem{tJ} Hirsch J. E., Phys. Rev. Lett. \textbf{54}, 1317 (1985);  Spalek J., Phys. Rev B \textbf{37}, 533 (1988);
 Gros C.,  Joynt R. and Rice  T. M., Phys. Rev. B \textbf{36}, 381 (1987).



\bibitem{kiv}  Emery V. J.,  Kivelson S. A. and  Lin H. Q., Phys. Rev. Lett. \textbf{64}, 475 (1990).

\bibitem{hague}  Hague J. P.,  Kornilovitch P. E.,  Samson J., and  Alexandrov A.
S., Phys. Rev. Lett. \textbf{98}, 037002 (2007).

\bibitem{alekabmot}  Alexandrov A. S.,  Kabanov V. V. and Mott  N. F., Phys. Rev. Lett. \textbf{77}, 4796
(1996).
\bibitem{aleran}  Alexandrov A. S. and  Ranninger J., Phys. Rev. B \textbf{23}%
, 1796 (1981); Phys. Rev. B \textbf{24}, 1164 (1981).

\bibitem{alemot}
 Alexandrov, A. S. and  Mott, N. F. {\em Polarons and Bipolarons}, p. 144
(World Scientific, Singapore, 1995).

\bibitem{Uemura}  Uemura, Y. J. {\em et. al},  Universal Correlations between
$T_{c}$ and $n_{s}/m^{*}$ (Carrier Density over Effective Mass) in
High-T$_{c}$ Cuprate Superconductors. \emph{Phys. Rev. Lett.} {\bf
62}, 2317-2320 (1989).



\bibitem{Will} Willemin, M.,   Rossel, C.,  Hofer, J.,  Keller, H., and
Revcolevschi, A. Anisotropy scaling close to the ab plane in
La$_{1.9}$Sr$_{0.1}$CuO$_{4}$ by torque magnetometry. \emph{Phys.
Rev. B} {\bf 59}, R717-720 (1999).


\bibitem{Basov} Basov, D. N. {\em et al.}, In-Plane Anisotropy
of the Penetration Depth in YBa$_{2}$Cu$_{3}$O$_{7-x}$ and
YBa$_{2}$Cu$_{4}$O$_{8}$ Superconductors. \emph{Phys.  Rev.  Lett.}
{\bf 92}, 598-601 (1995).

\bibitem{Chma} Chmaissem, O.,  Eckstein,  Y., and  Kuper, C. G. Structure
and a bond-valence-sum study of the 1-2-3 superconductors
(Ca$_{x}$La$_{1-x}$)(Ba$_{1.75-x}$La$_{0.25+x}$)Cu$_{3}$O$_{y}$ and
YBa$_{2}$Cu$_{3}$O$_{y}$. \emph{Phys. Rev. B} {\bf 63},
174510-174519 (2001).

\bibitem{Taj}S. Tajima, J. Schu¬tzmann,S. Miyamoto, I. Terasaki, Y. Sato, and R.
Hauff. Optical study of c-axis charge dynamics in
YBa$_{2}$Cu$_{3}$O$_{y}$: Carrier self-confinement in the normal and
the superconducting states. \emph{Phys. Rev. B} {\bf 55}, 6051-6060
(2001).


\end{thebibliography}
\end{document}